# Heat transfer in rough nanofilms and nanowires using Full Band Ab Initio Monte Carlo simulation

B. Davier[1], J. Larroque[1], P. Dollfus[1], L. Chaput[2], S. Volz[3], D. Lacroix[2], J. Saint-Martin[1]

[1] C2N UMR 9001, Univ. Paris-Sud-CNRS, Université Paris-Saclay, 91405 Orsay, France

[2] Université de Lorraine,CNRS, LEMTA,F-54000 Nancy, France

[3] LIMMS UMI 2820, The University of Tokyo-CNRS, 4-6-1 Komba, Meguro-ku, Tokyo 153-8505, Japan

## I. Abstract

The Boltzmann transport equation is one of the most relevant framework to study the heat transport at the nanoscale, beyond the diffusive regime and up to the micrometer-scale. In the general case of three-dimensional devices, the particle Monte Carlo approach of phonon transport is particularly powerful and convenient, and requires reasonable computational resources.

In this work, we propose an original and versatile particle Monte Carlo approach parametrized by using *ab-initio* data. Both the phonon dispersion and the phonon-phonon scattering rates have been computed by DFT calculation in the entire 3D Brillouin zone. To treat the phonon transport at rough interfaces, a combination of specular and diffuse reflections has been implemented in phase space.

Thermal transport has been investigated in nanowires and thin films made of cubic and hexagonal Silicon, including edge roughness, in terms of effective thermal conductivity, phonon band contributions and heat flux orientation. It is shown that the effective thermal conductivity in quasi-ballistic regime obtained from our Monte Carlo simulation cannot be accurately fitted by simple semi-analytical Matthiessen-like models and that spectral approaches are mandatory to get good results. Our Full Band approach shows that some phonon branches exhibiting a negative group velocity in some parts of the Brillouin zone may contribute negatively to the total thermal flux. Besides, the thermal flux clearly appears to be oriented along directions of high density of states. The resulting anisotropy of the heat flux is discussed together with the influence of rough interfaces.

## II. Introduction

The optimization of the thermoelectric conversion is an active subject of research. The main applications are related to energy harvesting for power supply autonomous systems as well as heat management in CPU cooling. Yet, common and efficient thermoelectric materials such as Bismuth Telluride, Lead Telluride, etc., often rely on unfortunately rare and toxic compounds. Their replacement by Silicon and Germanium which are more abundant and widely used in microelectronics would be appealing if their naturally poor thermoelectric properties could be significantly improved, especially close to room temperature.

The thermoelectric efficiency of a material is characterized by a unitless thermoelectric figure of merit $ZT = S^2 \sigma T / \kappa$ that depends on the electrical conductivity $\sigma$, the Seebeck coefficient S and the thermal conductivity $\kappa$. To improve the conversion efficiency, $ZT$ has to be increased. Consequently, $\kappa$ must be reduced while $\sigma$ must be preserved as far as possible. As these two parameters are strongly inter-dependent in common bulk materials, ZT optimization remained very limited for decades[1].

However, nanotechnologies provide new routes to optimize the thermoelectric conversion[2] For instance, nanostructures with a characteristic width $W$ both larger than the mean free path of electrons $l_{mfp}^e$ and smaller than the mean free path of phonons $l_{mfp}^{ph}$ ($l_{mfp}^e < W < l_{mfp}^{ph}$) can be used to specifically control and limit the effective phonon mean free path. This way, since heat transfer in non-degenerate semiconductors is mainly due to phonons, higher ratios $\sigma/\kappa$ can be achieved and finally ZT can be significantly higher in such nanomaterials than in their bulk counterpart. For instance, experimental measurements in Silicon nanowires of appropriate diameters[3] i.e. on the order of 100 nm, have demonstrated a large reduction of the effective thermal conductivity[4–6], down to

2 orders of magnitude lower than bulk conductivity. Due to limited electrical conductivity reduction, *ZT* values higher than 1 were demonstrated[7].

To estimate numerically the thermal properties, *ab-initio* DFT (Density Functional Theory) simulations have been used for studying properties of bulk materials[8] or nanostructures of some nanometers size[9,10]. However, DFT simulation of nanostructures larger than $l^e_{mfp}$, i.e. a few tens of nanometers in common materials, becomes very expensive in terms of computational resources. Molecular Dynamics (MD) simulations have also been widely applied for modeling nanowires made of different crystalline structures[11] and of diameter up to 20 nm[12,13]. These atomistic simulations can describe the effect of rough interfaces in real space[14,15] and thus the resulting reflection of mechanical waves. However, MD relies on empirical inter-atomic potentials that must be properly adjusted and, in principle, this classical approach of transport is accurate only at high temperature, higher than the Debye's temperature, that is 640 K in Si[16].

Complementarily, methods based on the resolution of the Boltzmann Transport Equation (BTE) for phonons are relevant for larger dimensions and over the full temperature range. The BTE can be solved analytically when simple phonon dispersions and scattering terms are considered as in pioneering works of Callaway[17] and Holland[18]. Mingo and co-workers[19] have used a full-band dispersion relation and Kazan[20] a more complex model for rough boundaries.

In 1D systems, the BTE can be solved by a direct approach[21], but for 3D problems, a stochastic particle Monte Carlo method[22] is much more efficient and it is able to include complex scattering terms[23]. This versatile approach can solve accurately the BTE much beyond the linear approximation, and in complex geometries. It has been used in bulk material configurations[24–27], porous nanofilms[28,29] and nanowires with regular shapes[30,31]. The reflection at external rough interfaces can be implemented considering specular reflections at boundary with realistic shape in real space i.e. a saw-tooth shape[32] or a random surface[33]. Other models consider a specific scattering term related to a diffusive reflection at the interface that tends to randomize the propagation direction of diffused phonons. Casimir has linked the thermal conductivity to the width of structures[34]. More recently, a probability of specular reflection[35,36] and a characteristic length of the diffusive reflections in the case of ultra-thin wires[37] were defined. In Soffer's work[38], this probability depends on incident wave vector of phonons plus two empirical parameters: the surface roughness standard deviation and its tangential correlation.

In contrast to the *ab-initio* approach, the accuracy of semi-empirical methods such as BTE or MD depends on the choice of their input parameters (inter-atomic potential, dispersion properties, phonon lifetimes, etc.). They are thus not convenient to study a new material for which such parameters are unknown. Recently, a methodology for solving the BTE in thin films without any adjustable parameter has been presented in Ref. **39**. It makes use of BTE parameters, i.e. full-band dispersion and full-band three-phonon scatterings rates, preliminarily extracted from DFT calculation.

This work extends the approach developed in Ref. 39 by the implementation of phonon reflection at rough interfaces in the full-band Monte Carlo simulation parametrized by *ab-initio* calculations. This advanced simulation method is used here to compute the thermal resistance in Si nanofilms and nanowires of several crystalline orientations and phases (cubic and hexagonal). The anisotropy of the heat transfer and the transition from ballistic to diffusive transport regimes are carefully investigated. Additionally, the relevance of semi-analytical models in such systems is discussed.

In Section III, our *ab-initio* parametrized MC simulator, the rough interface model as well as the relevant semi-analytical formalisms are detailed. In Section IV, the computed thermal conductivities of nanowires and thin films in both cross-plane and in-plane configurations are investigated.

# III. Models

### III.1 *Ab-initio* based material parameters

To solve the transport equation for any kind of particles in a solid-state system, the prior knowledge of both the energy dispersion of particles and their scattering rates is required. For phonon transport, many previous works assumed a simple isotropic dispersion relationship[24,26,27,30,40]. However, this approximation cannot reproduce accurately both the heat capacity and the heat conductivity[41] using a single normalization parameter. In the present work, the energy dispersion, the phonon velocity as well as the phonon scattering rates in all directions have been considered. Indeed, all states belonging to the Brillouin Zone (BZ) have been considered in our "Full-Band" description.

To determine all these "Full-Band" material parameters, a powerful *ab-initio* method was used, which is relevant to investigate accurately the phonon properties of a large range of materials. They were calculated by using DFT simulation as detailed in Ref. 39,42.

The first BZ was discretized in $N$ wave vectors, with $N = 31 \times 31 \times 31 = 29791$ and $N = 31 \times 31 \times 19 = 18259$ for cubic Silicon (Si3C) and hexagonal Silicon (Si2H), respectively. Si3C and Si2H have 6 and 12 phonon modes, respectively. The angular frequency, the group velocity and the phonon-phonon scattering rates were calculated for each discrete state characterized by a couple of a wave vector $\vec{q}$ and a mode $m$.

The only phonon scattering rates computed here by DFT are those related to the intrinsic phonon-phonon scattering mechanisms that are dominant in bulk materials. These rates $\lambda$ were calculated via the finite displacement method detailed in supplementary materials of Ref. 39,42, for 101 temperatures ranging between 0 and 1000 K. Throughout the simulation, the values corresponding to intermediate temperatures were interpolated by using a cubic spline method.

The discrete nature of this description must be taken into account, in particular to define the iso-energy states used for instance to select the final state after any scattering event. The strict conservation of energy or frequency in this case must be relaxed to frequency variations smaller than a discretization step $\Delta\omega$. In our simulation $\Delta\omega$ was defined as $\Delta\omega = \omega_{max}/128$ where $\omega_{max}$ is the maximum phonon frequency in the material (this value of $\Delta\omega$ leads to negligible average energy loss as shown in supplementary materials[43]). The resulting iso-energy curves of the first phonon mode in both cubic and hexagonal Silicon are plotted in Fig. 1. Parts (a) and (b) represent the BZ in face centered cubic and hexagonal lattices (for Si3C and Si2H, respectively) and their high symmetry points of interests (from Ref. [44]). Parts (c), (d), (e), and (f) map the angular frequency in the main cutting planes of the BZ. In the cutting plane $\Gamma X L$ of cubic silicon in (c), iso-energy curves are far from being circular (spherical in 3D) as in isotropic materials in the whole frequency range. In particular, the anisotropy between the L and U points is strong. In the hexagonal phase in (d), (e), and (f), an isotropic behavior is nearly achieved in all planes ($K\Gamma M$, $M\Gamma A$, and $K\Gamma A$, respectively) but only at low frequencies. In contrast, far from the $\Gamma$ point the anisotropy becomes strong, in particular between the M and K points.

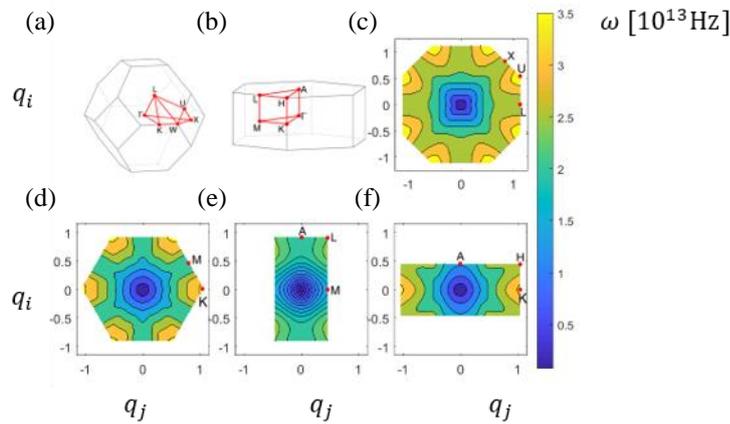

FIG. 1. Cartography of the angular frequency $\omega$ in the BZ. Wave vectors $q_{i,j}$ are in $\left[10^{10}\frac{2\pi}{m}\right]$. Schema of the BZ in (a) Si3C, (b) Si2H. Isoenergies in (c) the (110) plane of Si3C, (d),(e),(f) in planes of Si2H.

### III.2 Investigated devices

Two different types of nanostructures, i.e. nanofilms and nanowires, were investigated. In nanofilms, the external interfaces, separated by a finite distance, are called the top and bottom interfaces. Both the in-plane and cross-plane configurations were considered, depending on whether the thermal flux is parallel or perpendicular to the interfaces, respectively.

As schematized in Fig. 2, to implement numerically the devices, a cubic mesh was used. A material (of arbitrary crystal orientation) is assigned to each cell. All cells are aligned and located in between a cold thermostat (blue plane) and a hot one (red plane). A face of a cell can be either transparent (when the adjacent cell is made of the same material), specular (no adjacent cell) or diffusive (green planes). The incident angle of a particle colliding with a specular face is equal to the reflected one. This reflected angle is of course different when a diffusive face is involved and must be selected carefully, as explained later. To mimic infinite dimensions, as needed in cross-plane configuration for all directions except the transport one, specular reflections are implemented at both opposite boundaries. In this study, the heat transport is along the X-axis. The three devices are defined by the type of boundaries, as shown in Fig. 2, that is

(a) CP nanofilm, i.e. orientated in the cross-plane direction, with only specular boundaries;

(b) IP nanofilm, i.e. orientated along in-plane direction with specular (XZ planes) and diffusive (XY planes) opposite boundaries (colored in green in Fig. 2);

(c) rough nanowire, with only diffusive boundaries.

These devices are parameterized by their length $L$ (distance between thermostats along X axis) and their width $W$ (along the $Z$ axis). In this study, only nanowires with a square cross section are considered (i.e. with the height along Y direction equal to $W$).

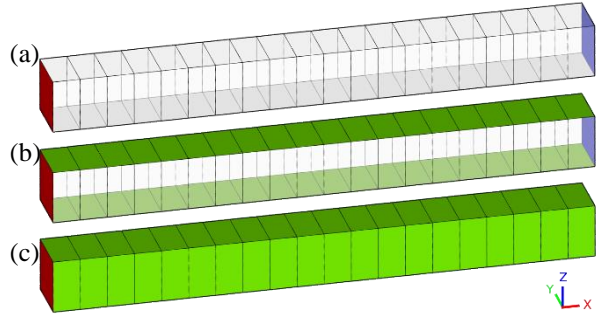

FIG. 2. Schema of simulated nanostructures: (a) CP nanofilm in cross-plane configuration, (b) IP nanofilm in in-plane configuration, (c) rough nanowire. Red/blue faces for hot/cold thermostats. Transparent/green faces for specular/rough boundaries.

### III.3 Monte Carlo simulation

The Boltzmann Transport Equation (BTE) describes the time evolution of the phonon distribution function in phase space $f_j(\vec{r}, \vec{q}, t)$, where $j$ is the phonon state, $\vec{r}$ the position in real space, $\vec{q}$ the wave-vector and $t$ the time. Its general expression is given by:

$$\frac{\partial f_j}{\partial t} + \vec{v}_j \cdot \vec{\nabla} f_j = \frac{\partial f_j}{\partial t}\bigg|_{interaction}, \qquad (1)$$

where $\vec{v}_j$ is the phonon group velocity ($\vec{v}_j = \frac{\partial \omega_j}{\partial \vec{q}}$, $\omega_j$ is the phonon angular frequency) and $\frac{\partial f_j}{\partial t}\bigg|_{interaction}$ is the rate of scattering from state $j$ to any another state.

In order to solve the BTE, we use the particle Monte Carlo method for phonon transport[25]. In this stochastic approach, the trajectories of a large number of particles are randomly selected. These trajectories are described as a succession of free flights (linear trajectories in real space without any change in the reciprocal space) separated by instantaneous scattering events. The scattering mechanism ending a free-flight can be either a phonon-phonon scattering or a collision with the device boundary that modifies only the wave vector coordinates. The initial state of particles, the duration of each free flight, the type and the effect of each scattering event are chosen randomly according to all relevant scattering rates. Finally, the phonon distribution is reconstructed from summing over all the particles $k$ belonging to the mode $m$ at a given time:

$$f_m(\vec{r}, \vec{q}, t) = \sum_{\text{particles } k} \delta[\vec{q} - \vec{q}_k(t)] \, \delta[\vec{r} - \vec{r}_k(t)] \qquad (2)$$

The wave vector-dependent relaxation time approximation is used for the phonon-phonon scattering rate, as follows:

$$\left.\frac{\partial f_m}{\partial t}\right|_{interaction} = -\left(f_m(\vec{r},\vec{q},t) - f_m^{eq}(\vec{q},T_c)\right)\lambda_m(\vec{q},T_c) \quad (3)$$

Thus, under the effect of particle scattering, the system tends "naturally" to recover its (Bose-Einstein) equilibrium distribution at the local temperature $T_c$.

In the following sub-section, the overall Monte Carlo algorithm is presented. Then the different kinds of external boundaries are described. Finally, the implementation and the post-processing methods are detailed.

### III.3.1 Effective temperature

The temperature in each device cell must be updated during the simulation, as the phonon-phonon scattering rates are temperature-dependent. The local effective temperature $T_c$ is related to the energy density $E_V$ through a relationship at equilibrium that has been preliminary tabulated according to Bose-Einstein statistics for each material. Then, during the simulation the energy density $E_{V,c}$ in each cell (that it is proportional to the local particle density) is periodically updated and thus the local temperature is extracted through energy inversion: $T_c = E_V^{-1}(E_{V,c})$. The equation for the energy and the resulting energy-density relationship are shown in Supplementary Materials[43].

### III.3.2 Definition of simulated particles

To reduce the particle number and thus the computational resources, in our model each simulated particle represents a packet of $N_\omega$ phonons with a frequency $\omega$. Considering phonon packets with a constant number of phonons $N_\omega = N$ (and thus particles of different energies $E_p = N \times \hbar\omega$ according to $\omega$) would make difficult the treatment of elastic interactions[25]. Thus, the number of phonons in a packet $N_\omega$ was chosen depending on the frequency $\omega$ and it was tuned to guaranty the same total energy $E_p$ ($= N_\omega \times \hbar\omega$) for all simulated particles whatever their frequency. This input parameter $E_p$ defines the energy resolution of the simulation.

Since we are mainly interested in the deviation of phonon distributions from equilibrium state, we simulated only the excess particles with respect to the equilibrium distribution at a reference temperature $T^0$ close to and usually below the actual temperature in the device. Every simulated particle gets a sign $s$ equal to +1 or -1 to represent an excess or a missing particle, respectively, with respect to the reference distribution.

For instance, the density of particles in a state $j$ in equilibrium at temperature $T$ is:

$$n_j^{eq}(T) = \frac{\hbar\omega_j}{E_p}\left(f_{B-E}(\omega_j,T) - f_{B-E}(\omega_j,T^0)\right)\frac{V_j}{(2\pi)^3}, \quad (4)$$

where $f_{B-E}$ is the Bose-Einstein distribution, and $V_j$ the reciprocal volume of the state.

This technique reduces both the simulation time and the numerical noise (see Supplementary Materials for more information[43]). All simulations presented in this work were based on this approach called "energy-based variance-reduced method" by Péraud and co-workers[25].

### III.3.3 Particle Monte Carlo algorithm

After initialization (see III.3.4), two nested loops are performed, i.e. one over time and the other one over particles. At every time step, particles are injected from the thermostats (see III.3.5). Then, the displacement of each particle is computed and its coordinates ($\vec{r}, \vec{q}$) are updated (see III.3.6). During a time step, the events that can interrupt the free flight are either a collision with a cell boundary or a phonon-phonon scattering event (see III.3.7 and III.3.8). A flowchart diagram of this algorithm is included in Supplementary Materials[43].

### III.3.4 Initial conditions

In a cell of volume $V_c$, the initial number of particles $N_{m,initial}$ in a mode $m$ is numerically calculated by using the equilibrium density $n^{eq}(\omega_j, T_c)$ defined in (4) and summing over all ($m$-mode) states as follows:

$$N_{m,initial} = V_c \sum_{\substack{\text{state } j \\ \text{of mode } m}} |n^{eq}(\omega_j, T_c)|, \quad (5)$$

where $E_p$ and $T_c$ are the energy of a simulated particle and the local temperature, respectively.

For each initial particle, the angular frequency is selected according to a distribution proportional to the volume of each iso-energy state defined in III.1. Then, its wave-vector is randomly and uniformly selected among the iso-energy states. The position in the cell is selected according to a uniform distribution. Finally, the sign of the particle (see III.3.2) is positive if the local temperature $T_c$ is higher than the reference temperature $T^0$, otherwise the sign is negative.

### III.3.5 Thermostats

Thermostats inject a constant flux of particles, according to their temperature. The number of particles injected during a time step $\delta t$ through a surface is:

$$N_{thermostat}(T) = \sum_{\substack{state\ j \\ \vec{v}_j \cdot \vec{n}_\perp > 0}} \vec{v}_j \cdot \vec{n}_\perp\ A\ dt\ n_j^{eq}(T), \qquad (6)$$

where $\vec{n}_\perp$ is the unit vector normal to the thermostated face and $A$ is its area.

As each particle keeps track of its own simulated time, we can initialize all of them at the beginning of the time step and make them behave as if they were injected continuously.

Another possible method to simulate thermostats is to have an additional cell behind the thermostated face. Every face of that "blackbody" cell is specular, so that its phonon distribution is constant over time. When a particle collides with the face connected to the device, a duplicate particle is transmitted. We implemented both methods and confirmed they produce the same heat flux. The direct injection at surfaces was chosen, as it is less computationally intensive.

### III.3.6 Time of free flight and transport

The free flight corresponds to the movement of a particle between two scattering events. The scattering events can be separated in standard (bulk material) scattering mechanisms such as phonon-phonon scattering and phonon-boundary scattering.

For each free flight, the time before the next standard scattering event $t_{scattering}$ is randomly selected according to the following formula:

$$t_{scattering} = -\frac{\ln(n_{random})}{\lambda_j(T_c)}, \qquad (7)$$

where $n_{random}$ is a uniform random number in $]0; 1]$ and $\lambda_j(T_c)$ is the total scattering rate from a phonon in a state $j$ at a local temperature $T_c$. These scattering rates $\lambda_j(T_c)$ are assumed to be constant during a time step. In the general case, $\lambda_j(T_c)$ is the sum of the scattering rates corresponding to the different scattering mechanisms, assumed to be independent. In this work, only the phonon-phonon scattering rates contribute to $\lambda_j(T_c)$. They are calculated by an *ab-initio* approach as previously mentioned. This method was first developed for electron transport[45].

The time before the next boundary collision is simply derived from the distance $\vec{d}$ between the particle and the boundaries along its transport direction. As the phonon velocity is constant during a free flight, we have

$$t_{boundary} = \min\left(\frac{d^x}{v_j^x}, \frac{d^y}{v_j^y}, \frac{d^z}{v_j^z}\right) \qquad (8)$$

The free flight duration $t_{ff}$ for the particle is thus limited by the first event that occurs, i.e.

$$t_{ff} = \min(t_{remaining}, t_{boundary}, t_{scattering}), \qquad (9)$$

where $t_{remaining}$ is the remaining time before the end of the *i-th* time step $\delta t$ for the particle. The interruption of a free flight by the end of a time step has no impact on the other scattering rates since they are Poissonian processes.

$$t_{remaining} = (t_i + \delta t) - t_{current,k} \qquad (10)$$

At the end of the free flight, the position in real space $\vec{r}_k$ of the particle $k$ with a velocity $\vec{v}_j$ is updated according to:

$$\vec{r}_k(t_{current,k} + t_{ff}) = \vec{r}_k(t_{current,k}) + \vec{v}_j(t_{current,k})\ t_{ff} \qquad (11)$$

The update of the wave vector depends on the event that interrupts the free flight (phonon-phonon scattering, or boundary reflection). Finally, the time counter of the particle is updated. New free flights are selected for that particle until it has been simulated for the full timestep.

$$t_{current,k} = t_{current,k} + t_{ff} \qquad (12)$$

### III.3.7 Phonon-phonon scattering

If a phonon-phonon scattering event occurs at the end of the free flight (i.e. $t_{ff} = t_{scattering}$) a new phonon state has to be selected. The memory of the initial state is lost[23], and the three-phonon processes are simplified into two-phonon processes: the scattered phonon is destroyed and the new one is randomly selected in order to recover an equilibrium distribution as proposed in the work of Lacroix et al.[31].

The probability of selecting a new state $j$ is proportional to the equilibrium density of particles weighted by the interaction rate of that state, i.e.

$$P_j \propto \lambda_j(T_c)\, n_j^{eq}(T_c) \qquad (13)$$

### III.3.8 Specular and diffusive boundaries

When a particle collides with external boundaries, two kinds of reflection may occur, i.e. either a specular or a diffusive one.

At smooth boundaries, the particle reflection is always specular, i.e. the wave-vector component normal to the surface boundary of the reflected particle ($q_{\perp 0}'$) is the reverse of the incident one ($q_{\perp 0} = -q_{\perp 0}'$). It should be mentioned that the implementation of specular reflection is not easy within a full-band description when orientation of the boundary does not correspond to a high symmetry plane of the crystal. As such specular reflections have no impact on the thermal flux parallel to the interface, they are used in our simulations to emulate semi-infinite boundaries.

In the case of a collision with a rough boundary, the particle has a given probability to undergo a diffusive reflection that randomizes the final wave vector instead of specular reflection. In this work, the following expression of probability of specular reflections has been used[38]:

$$p_{specular}(|\vec{q_0}|, \cos(\theta_0)) = \exp[-(2\cos(\theta_0)\, \Delta\, |\vec{q_0}|)^2] \qquad (14)$$

where $\vec{q}_0$ is the incident wave-vector, $\Delta$ is an empirical surface roughness parameter and $\theta_0$ is the incident angle defined as:

$$\cos(\theta_0) = \frac{v_\perp}{|\vec{v_j}|} \qquad (15)$$

where $v_\perp$ is the component perpendicular to the boundary interface of the velocity $\vec{v_j}$ of the incident phonon in state $j$. This definition is specific to the full-band dispersion in which the velocity and wave-vector of a given phonon state are not necessarily collinear.

As a diffusive reflection conserves the energy, the final state $j'$ is selected among the states belonging to the same iso-energy surface and with a good orientation of the final velocity (i.e. $v_\perp . v_\perp' < 0$). The probability of each reflected state is weighted by two factors: the complementary probability of a specular reflection for a reflected state $j'$ with an angle $\theta_r$, and the related normal component of group velocity as follows:

$$p_{j,diffusive}(|\vec{q}_{j'}|, |\vec{v}_{j'}|, \cos(\theta_r)) \propto \left(1 - p_{specular}(|\vec{q}_{j'}|, \cos(\theta_r))\right) |\vec{v}_{j'}| \cos(\theta_r) \qquad (16)$$

This angular distribution of the final state allows conserving the angular distribution of the heat flux at equilibrium temperature, which leads to a net flux along the normal direction equal to zero at the steady state (as in the Lambert's law) across simulated adiabatic interfaces. This prevents an unphysical phonon accumulation from occurring near the rough boundary.

### III.3.9 Post-processing

During the simulation, at each time step and in each cell, the local temperature $T_c$ is calculated as described in III.3.1 as well as the local thermal heat flux density $J_c$ [W m$^{-2}$] by summing the contribution of all particles.

The thermal conductivity of the simulated device is then calculated from the average heat flux density $\vec{J}_c$ along the transport direction $\vec{n}$ by

$$\kappa_{simulation} = \frac{\vec{j}_c \cdot \vec{n}}{\Delta T} L \qquad (17)$$

with the temperature difference $\Delta T$ and length $L$ between the thermostats. The confidence interval at 95% was calculated for all simulations and was found smaller than 1 W m$^{-1}$ K$^{-1}$ except in long devices ($L = 100\mu m$) for which the precision was reduced to achieve reasonable simulation times.

### III.3.10 Simulation parameters

We used several criteria to select the simulation parameters. First, we estimate the thermal relaxation time of the device from the appropriate diffusive thermal conductivity $\kappa$ (see III.4), the volumetric heat capacity $\frac{\partial E_V}{\partial T}$, and the distance $L$ between thermostats:

$$\tau = \frac{\partial E_V}{\partial T} \frac{L^2}{\kappa} \qquad (18)$$

Then, the time step duration $\delta t$ is set to $\tau/20$. The temperature and heat flux are averaged every 5 $\tau$.

Finally, to choose a relevant particle energy $E_p$, the temperature fluctuation due to one particle displacement in a volume $V_c$ is chosen to remain below $\delta T$ (typically equal to 0.01K) leading to:

$$E_P = \delta T \frac{\partial E_V}{\partial T} V_c \qquad (19)$$

These criteria result in an average number of 20 000 coexisting particles during the simulations. For a typical device of size 1 µm × 100 nm × 100 nm in Si3C, with a 4 K-temperature difference between thermostats and a reference temperature $T^0 = 295\,K$, our selected parameters were $E_p \approx 4\,10^{-18}$ J and $\delta t \approx 1$ ns, and the simulation lasted 20 minutes on a single thread. Besides, for devices of length ranging between 1 nm and 10 µm, the timestep $\delta t$ was scaled between about 0.1 ps and 100 ns, which affected the computation times accordingly.

### III.4 Semi-analytic models for thermal conductivity

In parallel with the numerical MC approach, the following ballistic and diffusive semi-analytic formula were used to estimate the conductivity for infinitely short and long CP nanofilms:

$$\kappa_{ballistic} = \frac{L}{V_s} \sum_{\text{state } j} \hbar \omega_j \left| \vec{v}_j \cdot \vec{n} \right| \frac{\partial f_{BE}}{\partial T}(\omega_j, T_{eq}), \qquad (20)$$

$$\kappa_{diffusive} = \frac{1}{V_s} \sum_{\text{state } j} \hbar \omega_j \left| \vec{v}_j \cdot \vec{n} \right|^2 \frac{1}{\lambda_j} \frac{\partial f_{BE}}{\partial T}(\omega_j, T_{eq}), \qquad (21)$$

where $\lambda_j$ is the phonon-phonon scattering rate for state $j$. Formula equivalent to Eq. (20) and Eq. (21) have been discussed for instance in Ref. 19. Eq. (20) comes from ballistic Landauer's formalism, i.e. without diffusive interaction and with a phonon transmission equal to 1. The resulting thermal conductivity is linearly dependent on the distance $L$ between thermostats. Eq. (21) is a solution of the linearized BTE within a diffusive transport approximation leading to a length-independent conductivity.

A first attempt to model the transition between both limits is based on a Matthiessen's rule that sums ballistic and phonon-phonon thermal resistances as follows:

$$\frac{1}{\kappa_{Matthiessen}} = \frac{1}{k_{diffusive}} + \frac{1}{\kappa_{ballistic}} \qquad (22)$$

In a more sophisticated approach, ballistic and phonon-phonon scattering rates are once again summed but their spectral dependences are considered. Indeed, the average distance over which a phonon moves in the transport direction $\vec{n}$ before colliding with a thermostat is

$$\bar{L} = \frac{L}{2} \qquad (23)$$

Similarly, in IP nanofilm or in nanowire, the average distance over which a phonon moves in a transverse direction $\vec{n}_\perp$ before colliding with a rough boundary is

$$\bar{W} = W \left( \frac{1}{1 - p_{specular}} - \frac{1}{2} \right), \qquad (24)$$

by considering the probability of specular reflection $p_{specular}$ from Eq. (14). This model is equivalent to the one proposed in Ref. 35.

For CP and IP nanofilms and for nanowires (NW), the total scattering terms for a state $j$ become:

$$\lambda_{j,NFCP} = \lambda_j + \frac{|\vec{v}_j \cdot \vec{n}|}{\bar{L}}, \tag{25}$$

$$\lambda_{j,NFIP} = \lambda_j + \frac{|\vec{v}_j \cdot \vec{n}|}{\bar{L}} + \frac{|\vec{v}_j \cdot \vec{n}_{\perp,1}|}{\bar{W}}, \tag{26}$$

$$\lambda_{j,NW} = \lambda_j + \frac{|\vec{v}_j \cdot \vec{n}|}{\bar{L}} + \frac{|\vec{v}_j \cdot \vec{n}_{\perp,1}|}{\bar{W}} + \frac{|\vec{v}_j \cdot \vec{n}_{\perp,2}|}{\bar{W}}, \tag{27}$$

Finally, the associated thermal conductivities $\kappa_{NFCP}$, $\kappa_{NFIP}$, and $\kappa_{NW}$ are calculated from Eq. (21) by replacing $\lambda_j$ by Eq. (25), (26), and (27) respectively.

## IV. Results

In this section, we investigate the dependence of thermal conductivity and the related heat flux on several parameters in both CP (cross-plane) and IP (in-plane) nanofilms and nanowires by using the code described in the previous section. We observe especially the transition between ballistic and diffusive transport regimes, the anisotropy of the thermal properties and the role of rough boundaries. All the following Monte Carlo simulations were performed at an average temperature of 300K. The temperature difference between the thermostats is of 4K, and the reference temperature is set at $T^0 = 295K$.

### IV.1 Cross-plane thermal conductivity

The first part of this section focuses on cross-plane thermal conductivity in CP nanofilms (see Fig. 2(a)) and the influence of their length $L$. The transport direction is along the $X$ axis. In the studied cases for cubic Silicon (Si3C), this real space direction corresponds to the [100] ($\Gamma X$) direction in the BZ. The thermal conductivity sometimes called effective thermal conductivity was computed via Eq. (17). Its evolution is plotted in Fig. 3 as a function of length $L$ in Si3C nanofilms. The Monte Carlo results (cross symbols) are compared with the different semi-analytical models presented in III.4.

In long devices, the thermal conductivity saturates to the diffusive value derived from Eq. (21) (dashed line). The calculated conductivity at the diffusive limit is $\kappa = 138\ [\text{Wm}^{-2}\text{k}^{-1}]$ at 300K. One should keep in mind that this value is derived without considering any phonon-impurity scattering mechanism (with just phonon-phonon scattering, see 3.1) and thus this value underestimates the experimental measurements of isotopically pure Silicon[46]. However, it is a good estimation of the thermal conductivity of natural Silicon at ambient temperature. Hence, no other scattering mechanism was taken into account in our study of bulk Silicon.

The thermal conductivity gradually changes from a diffusive regime of transport occurring in long nanofilms to a ballistic one in ultra-short films. We should note that for device length $L$ shorter than 10 nm, i.e. at the atomic scale, the considered phonon dispersion relation is not relevant and the indicated MC results are just a guide for the eyes illustrating the asymptotical behavior. Indeed, the linear behavior and the slope of the ballistic model (dotted line in Fig. 3) of Eq. (20) tends to be recovered in ultra-short films.

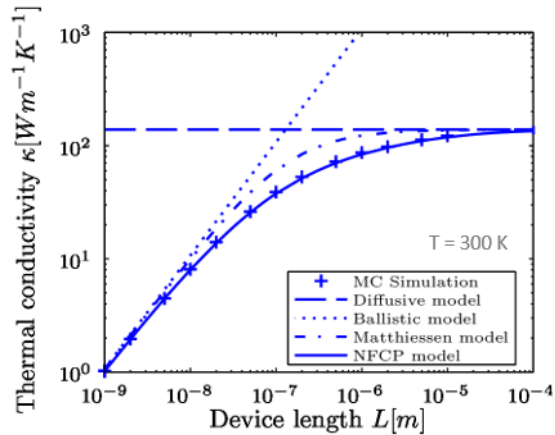

*FIG. 3. Cross-plane thermal conductivity κ as a function of length L for [100] Si3C nanofilms. Crosses: Monte Carlo simulations. Dotted, dashed, point-dashed and solid lines stand for models from Eq. (20), Eq. (21), Eq. (22) and Eq. (25), respectively.*

Between these limit cases, the Matthiessen model (Eq. (22) and point-dashed line in Fig. 3), which is commonly used in the literature, cannot properly capture the transition regime, i.e. the quasi-ballistic transport regime. It exhibits a difference up to 60% above the MC simulation results at $L = 200$ nm. However, the spectral NFCP model (Eq. (25) and solid line) is very close to MC results, with an underestimation lower than 4%.

The same simulations were performed on the [10-10] (called Si2Hx) and [0001] (Si2Hz) lattice orientations of the hexagonal phase of Si, and the results are plotted in Fig. 4 in green and red symbols, respectively. The same ballistic and diffusive asymptotic behaviors as in Si3C are observed. The conductivities in the hexagonal phase are always lower than in its cubic counterpart. There is also a significant anisotropy between the orientations of the hexagonal phase, as the diffusive thermal conductivity of Si2Hz is 26% lower than that of Si2Hx ($\kappa_{Si2Hx}^{diff} = 100$ Wm$^{-2}$K$^{-1}$, $\kappa_{Si2Hz}^{diff} = 74$ Wm$^{-2}$K$^{-1}$). Again, the spectral NFCP model (solid lines, see Eq. (25)) quite accurately reproduces the MC results in the quasi-ballistic region.

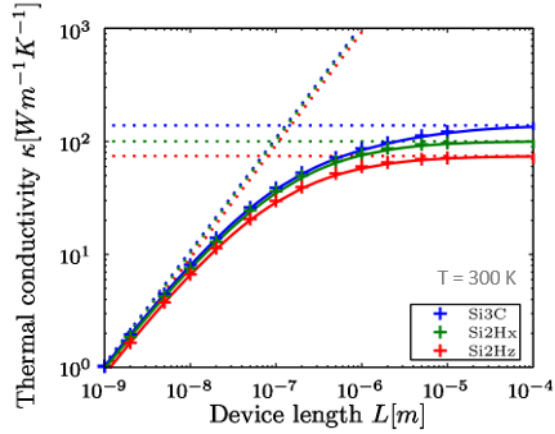

*FIG. 4. Cross-plane thermal conductivity κ as a function of length L for CP nanofilms of [100] Si3C, Si2Hx and Si2Hz. Crosses: Monte Carlo simulations. Dotted lines: ballistic and diffusive models. solid lines: NFCP model. T=300K*

## IV.2 Cross-plane heat flux distribution

We checked the spectral distribution of the particle energy and the heat flux of a 1 µm-thick film in Fig. 5(a) and (b), respectively. The different colors correspond to the contributions of the 6 modes of Si3C. As explained in III.3.3, the particles of a reference equilibrium distribution at $T^0$ (taken at 295K) are not considered. In Fig. 5(a), the energy distribution in the simulated device (cross symbols) is equal to the theoretical equilibrium at the local temperature (solid lines). Due to the phonon-phonon scattering processes, phonon energy distributions very close to equilibrium are observed in this 1 µm-long-device. Even with this energy distribution close to equilibrium, Fig. 5(b) shows that the cumulated heat flux is not perfectly described by the NFCP model (solid line), especially for the 3$^{rd}$ (LA) and 4$^{th}$ (LO) modes. This indicates the limit of the models based on linearized BTE (cf. Eq. (25)) to correctly account for the negative heat flux contributions present at high frequencies that are induced by some states that have group velocities $\vec{v_j}$ oriented in an opposite direction to their wave vector.

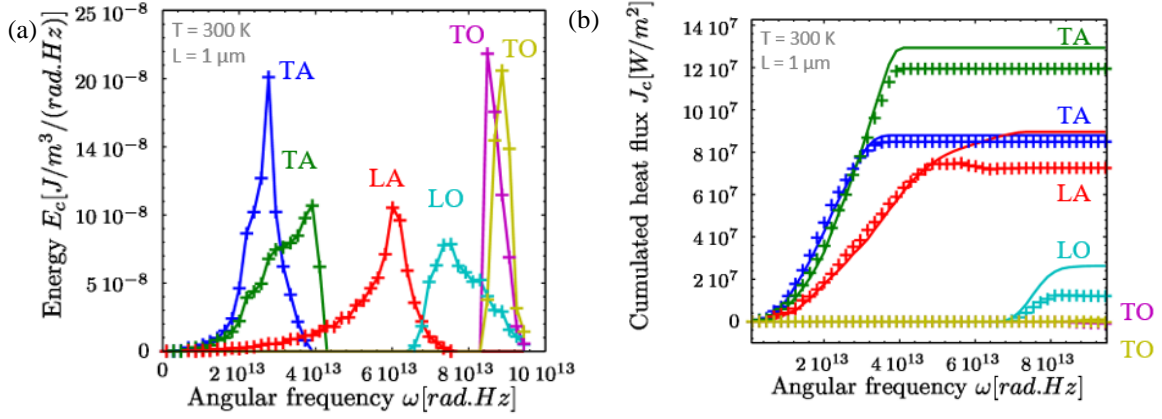

FIG. 5. [100] Si3C CP Nanofilm. Crosses: Monte Carlo simulations. (a) Angular frequency ω distribution of the energy density for each phonon mode. Solid lines: equilibirum distributions. (b) Cumulative cross-plane heat flux over frequencies. Solid line: NFCP model.

The contribution of each mode to the total heat flux is detailed in Table I and Table II for Si3C and Si2Hx, respectively. The modes are numbered by ascending energy values. In Si3C, about 96% of the heat flux is transported by acoustic modes in Si3C. Moreover, the 4$^{rd}$ (LO) mode carries around 5% of the total heat flux while the 5$^{th}$ and 6$^{th}$ (TO) modes have a negative net contribution, but small enough to be neglected as within the margin error. Their negative contribution reveals the contribution of states with a negative group velocity. In Si2H, around 90% of the heat is carried almost equally by the first 5$^{th}$ modes. For modes of higher energy than the 8$^{th}$ mode, the heat carried is fully negligible.

To estimate the anisotropy of the materials, we have analyzed in Fig. 6 the angular distribution of the heat flux. Figure 6(a) shows a schema defining the polar and azimuthal angles $\theta$ and $\phi$, relatively to the transport direction oriented along the $x$-axis. Then, the left ((b), (d), and (f)) and (right (c), (e), and (g)) columns are for Si3C and Si2Hx, respectively. The second row (b) and (c) shows the angular distribution of density of states (DOS). The angular DOS is related to the number of discrete states in the Brillouin zone which have a velocity's orientation within in a given solid angle (i.e. around the direction given by $\theta$ and $\phi$) and with a positive component of the velocity along X ($v_x$>0). Similarly, the third row (d) and (e) represents the angular distribution of the heat flux. This flux results from a sum over all trajectories of all particles during the simulation. These figures exhibit some noise especially along the directions of low DOS. The last row (f) and (g) displays the contribution of the same heat flux integrated over the polar angle $\theta$ and the azimuthal angle $\phi$. In an isotropic system, the angular distribution of the flux is expected to be a smooth cosine function. In both cubic and hexagonal phases, the angular distributions exhibit peaks revealing that the heat flux is mostly transported along specific orientations. In Si3C, while the angular DOS is highest in <110> directions, the heat flux is mainly transported along the [100] direction ($\theta = \pi/4$ and $\varphi=0$) and secondly along the <111> directions ($\theta = \pi/4$ and $\varphi = \pi/4$). In the anisotropic Si2Hx material, the transport is mainly focused in the hexagonal plane with $\theta = 0$. The main flux is along the [10-10] direction ($\varphi=0)$ and the two lateral peaks are along the <21-30> directions.

| Mode | Cross-plane | Rough Nanowires |
|---|---|---|
| 1 (TA) | 29.5 % | 37.0 % |
| 2 (TA) | 41.5 % | 39.9 % |
| 3 (LA) | 25.2 % | 18.9 % |

| Mode | Cross-plane | Rough Nanowires |
|---|---|---|
| 1 | 20.0 % | 18.4 % |
| 2 | 23.6 % | 20.7 % |
| 3 | 17.5 % | 20.0 % |
| 4 | 17.8 % | 15.7 % |
| 5 | 14.4 % | 14.5 % |
| 4 (LO) | 4.2 % | 4.8 % |
| 5 (TO) | -0.3 % | -0.4 % |
| 6 (TO) | -0.1 % | -0.2 % |

TABLE I. Relative heat flux contribution of Si3C phonon mode, in CP nanofilms and nanowires.

| | | |
|---|---|---|
| 6 | 1.2 % | 1.9 % |
| 7 | 1.8 % | 2.7 % |
| 8 | 2.8 % | 4.6 % |
| 9 | 0.2 % | 0.3 % |
| 10 | 0.2 % | 0.5 % |
| 11 | 0.3 % | 0.4 % |
| 12 | 0.1 % | 0.2 % |

TABLE II. Relative heat flux contribution of Si2Hx phonon mode, in CP nanofilms and in nanowires.

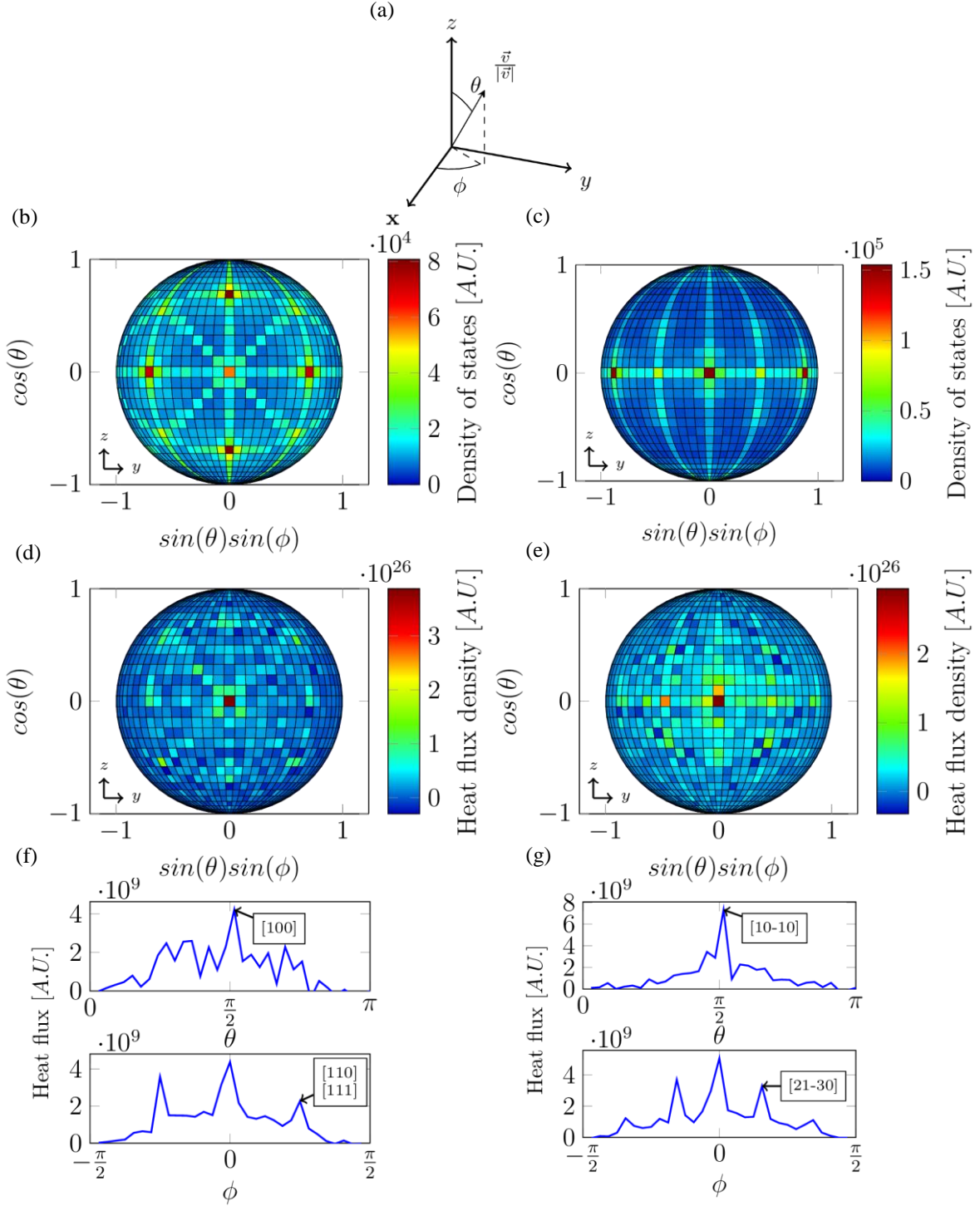

FIG. 6. CP Nanofilms at $T = 300K, L = 1\mu m$ of [100] Si3C: (b,d,f). [10-10] Si2H:(c,e,g). (a) "cross-plane"heat direction: x-axis, polar and azimuthal angles $\theta$ and $\phi$. (b) and (c) angular DOS. (d) and (e) angular distributions of the "cross-plane" heat flux. (f) and (g) integration over $\theta$ and $\phi$ of (d) and (e) , respectively.

## IV.3 Rough boundaries

In the previous part dedicated to the CP nanofilm, the reflection is specular at all external boundaries. This second part deals with the effect of rough boundaries, in the case of IP nanofilms (cf. Fig. 2(b)) and nanowires (cf. Fig. 2(c)) in which some reflections can be diffusive.

In FIG. 7, the evolution of thermal conductivity κ in an IP nanofilm and a nanowire is compared with that in CP nanofilm having no rough boundary. In Fig. 7 (a), the curve κ vs. length is plotted for $W$ = 100 nm where $W$ is either the width for nanofilms or the square cross section length for nanowires. Results provided by our MC simulation are indicated by crosses. In short devices in which the heat transport is ballistic, the thermal conductivities converge to the same values whatever the number of rough interfaces. This differs from the behavior in long devices as κ is reduced down to 87 and 69 W/m/K in IP nanofilms and nanowires, respectively, for a length of 100 µm. Thus, as expected, the conductivity reduction is directly related to the number of rough boundaries when the heat transport is diffusive. Besides, the semi-analytical models, indicated in solid lines, for IP nanofilms and nanowires do not fit MC results as well as in the case of CP nanofilm. For $L$=10 µm, the NFCP, NFIP, and NW models (cf. Eq. (25), (26) and (27), respectively) converge to a value 2%, 5% and 12% lower than the MC output, respectively (with $W = 100$ nm and $\Delta = 0.5$ nm).

In FIG. 7(b) the width $W$ dependence on thermal conductivity is plotted for 1 µm-long devices. IP nanofilms and nanowires show a similar behavior, i.e. starting from $\kappa$ at 16 and 6 $Wm^{-1}K^{-1}$ resp., increasing at about the same rate in the transition regime, and reaching 95% and 90% of the CP nanofilm conductivity at $W = 1$µm. The effect of rough boundaries thus becomes insignificant for wider devices. As all devices have a length of 1 µm, an intermediate heat transport regime occurs (see part 4.1). For devices wider than 1µm, the impact of rough boundaries is weak, and the cross-plane conductivity is recovered with a difference lower than 5%. For $W$ in the range of 10 nm to 200 nm the relationships between κ and W are quasi-exponential for both IP nanofilms and nanowires.

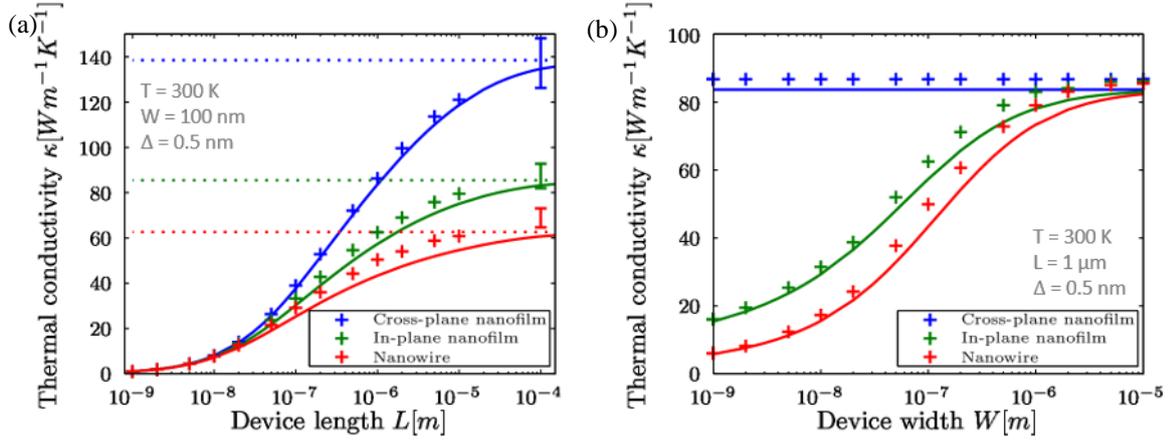

FIG. 7. Thermal conductivity κ as a function of (a) length L and (b) width W for IP and CP nanofilms and for nanowires of [100] Si3C. Solid lines: semi-analytical models. Dotted lines: long-device conductivity.

Table I and Table II detail the contribution of each mode to the heat flux in nanowires made of Si3C and Si2Hx, respectively. With respect to simulations with smooth boundaries (CP nanofilm), the introduction of rough boundaries in nanowires reduces the heat flux by 49% and 39% in Si3C and Si2Hx, respectively. However, the heat flux reduction is not uniform across the different modes of the material and the most impacted modes are the 2$^{nd}$ and 3$^{rd}$ ones in Si3C, and the 1$^{st}$, 2$^{nd}$ and 4$^{th}$ ones in Si2Hx.

The angular distribution of heat flux is plotted in Fig. 8 for a CP nanofilm and a nanowire made of Si2H oriented along the [100] direction and with a length $L = 1$ µm and width $W$ =100 nm. The lateral peaks of the heat flux corresponding to direction with a high density of states survive in rough films. However, they are strongly suppressed in the nanowire where the flux is much more focused along the transport direction [10-10] which is here also a main direction.

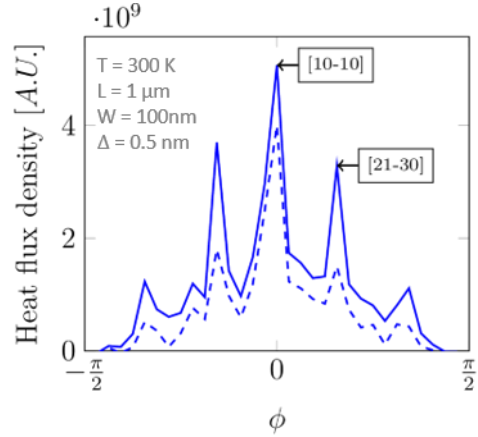

FIG. 8. Angular heat flux distribution as a function of azimuthal angle $\phi$, in [10-10] Si2Hx CP nanofilm (solid line) and nanowire (dashed line).

The evolution of the conductivity as a function of the surface roughness empirical parameter $\Delta$ is plotted in Fig. 9 for a 1 µm long nanofilm with a width $W = 100$ nm. Two plateaus can be observed. For ultra-small values of $\Delta$ lower than 0.1 nm, the diffusive reflections are negligible and then the cross-plane thermal conductivity is recovered. For value of $\Delta$ higher than 1 nm, the conductivities of nanowires and nanofilms with in-plane configuration reach their minimum. This minimum is related to a fully diffusive regime in which all phonon reflections at the external boundaries are diffusive. Besides, these Monte Carlo trends are reproduced by the relevant semi-analytical models i.e. NFIP (Eq. 26) and NW (Eq. 27). Nevertheless, these models tend to systematically underestimate the MC results.

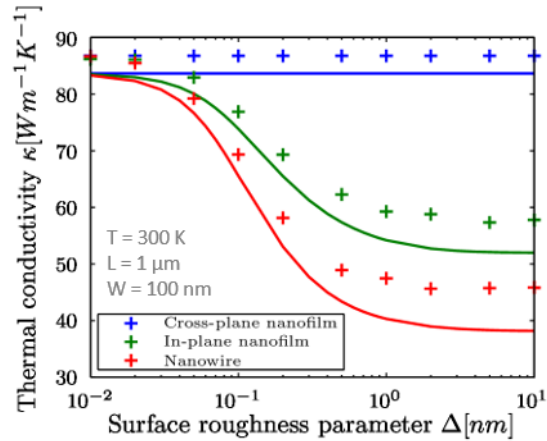

FIG. 9. Thermal conductivity $\kappa$ as a function of the surface roughness parameter $\Delta$ for CP nanofilms, IP nanofilms and nanowires of [100] Si3C. Crosses: MC resutls. Solid lines: semi-analytical models.

In order to mimic very rough external boundaries, a model called "fully diffusive" has been implemented, in which the probability of having a specular reflection $p_{specular}$ is always zero (this is equivalent to Soffer 's model with $\Delta > 1$ nm, see Fig. 9). To benchmark the Soffer's and fully diffusive boundary models, the two resulting conductivities in nanowires are plotted in Fig. 10 as a function of the length $L$. A significant discrepancy between the two models can be observed only for $L$ higher than 500 nm. This indicates that phonons have an average mean free path above this length in rough Si3C nanowires.

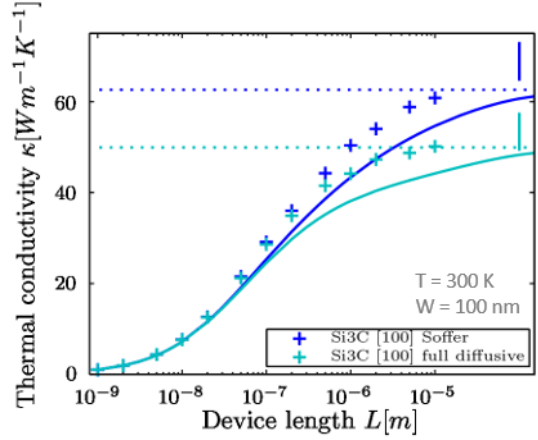

*FIG. 10. Thermal conductivity κ as a function of device length L for [100] Si3C nanowires by using Soffer's and fully diffusive boundary models. Crosses: MC resutls. Solid lines: semi-analytical models.*

Several experimental measurements of thermal conductivity in IP nanofilms and nanowires have been reported in the literature. We compared works from Ju[47], Liu[48] and Li[4] with results from our Monte Carlo simulation code in Fig. 11. In all these measurements, the device length $L$ is about 1 µm. In the case of IP nanofilms, the simulated thermal conductivities fit the experimental data of Liu[48] and slightly underestimate the conductivity from Ju[47]. For nanowires, our results are close to Li's ones and the thermal conductivity relationship with the width are relatively well reproduced. However, even our fully diffusive model overestimates the experimental value.

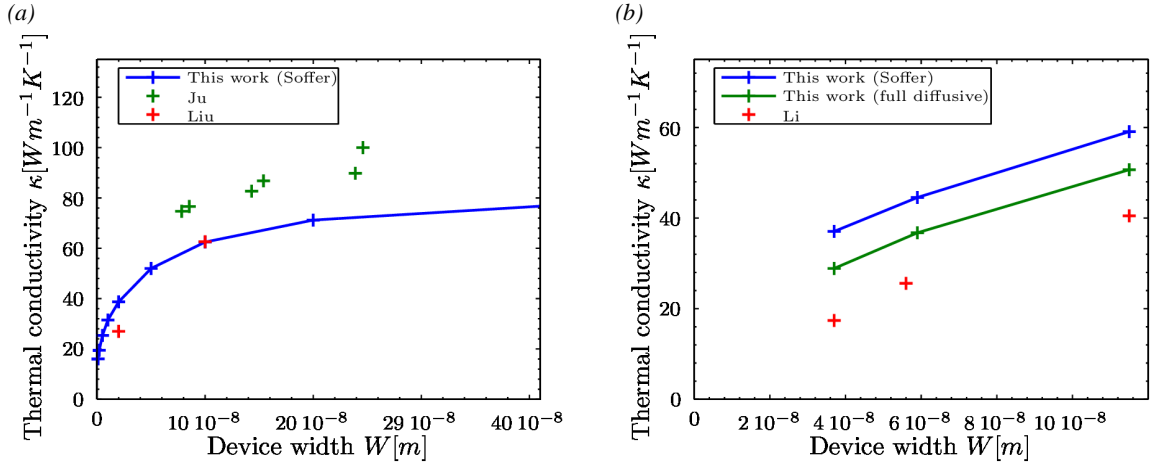

*FIG. 11. Thermal conductivity κ as a function of the width W in Si3C devices. Our MC results (crosses with solid lines) vs. experimental measurements. $T = 300K$, $L = 1µm$, $\Delta = 0.5nm$. (a) IP Nanofilms. Exp: Ju[47], Liu[48] (b) Nanowires. Exp: Li[4].*

## IV.4 Crystalline orientations

Even if only one crystalline orientation has been investigated for each crystalline phase in the previous sections our full-band approach allows *a priori* the study of arbitrary orientations. However, for our implementation of a specular reflection (although it is the standard one), the existence of a final state with a wave vector having a negative perpendicular component of the incident phonon ($q_{\perp 0} = -q'_{\perp 0}$) is mandatory. This condition requires that external faces of the device are aligned with a high symmetry plane of the crystal. This limitation does not apply in the fully diffusive model since the final state is randomly chosen among all the available final iso-energy states.

As this fully diffusive model has been previously shown to be relevant to study experimental nanowires (cf. Fig. 11 (b)), thermal conductivities have been simulated in Si3C nanowires for several crystal orientations. In Fig. 12, the thermal conductivity as a function of device length is plotted for [100], [110] and [111] lattice orientations by using the fully diffusive model. While the conductivities are similar in all directions for devices smaller than a few µm, at higher length different values are achieved, revealing some anisotropy in the heat transport. The long device limits along the [111] and [110] directions are 6% and 15% lower than the limit for the [100] direction, respectively. Once again, even if the semi-analytical models appear quantitatively disappointing in long devices when the fully diffusive approximation is used, they are able to capture the good trend of the orientation effect.

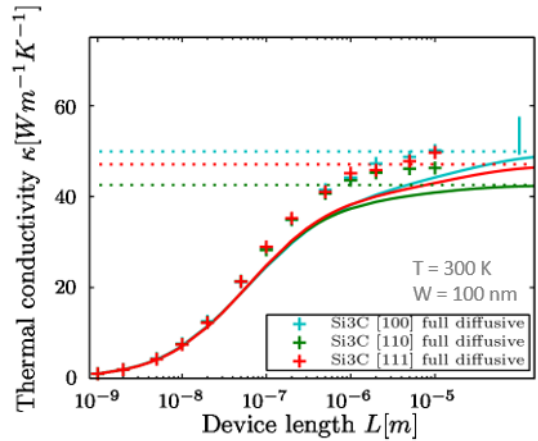

FIG. 12. Thermal conductivity κ as a function of device length L in Si3C nanowires for several crystalline orientations. Crosses: Fully diffusive MC results. Solid line: semi-analtyical models.

# V. Conclusion

A Full-band particle Monte Carlo algorithm dedicated to the phonon transport have been presented. As the material parameters are extracted from *ab-initio* calculation, this code can be used to study a large class of materials or crystalline phases. The phonon reflections at rough interfaces is modeled by using a specific scattering term requiring only one empirical parameter.

This code was used to study the heat transport in nanostructures such as nanofilms and nanowires for two phases of silicon (cubic and hexagonal). Thermal conductivity of both material is in agreement with theory and measurements and it is lower for hexagonal phase as compared to the cubic one. It has been shown that even if spectral semi-analytical models can estimate satisfactorily the cross plane thermal conductivity in quasi-ballistic transport in nanofilm, they are disappointing when the rough interface are dominant as in the case of nanofilms with in-plane configuration or in nanowires. Besides, the contribution of each phonon branch to the heat flux is complex and highly anisotropic. This is also depicted by polar and azimuthal descriptions of heat fluxes within nanofilms and nanowires. For the latter, the introduction of rough boundaries impact mostly acoustic modes (transverse acoustic for Si3C) and tends to focus the heat flux in the main transport direction.

In future works, the presented versatile numerical method allowing detailed treatment of the phonon interactions will be used to study devices with a more complex geometry made of semi-transparent interfaces.

# VI. Acknowledgements

This work is supported by a public grant overseen by the French National Research Agency (ANR) as part of the "Investissements d'Avenir" program (Labex NanoSaclay, reference: ANR-10-LABX-0035).

# VII. Bibliography

[1] G. Chen, M.S. Dresselhaus, G. Dresselhaus, J.-P. Fleurial, and T. Caillat, Int. Mater. Rev. **48**, 45 (2003).
[2] L.D. Hicks and M.S. Dresselhaus, Phys. Rev. B **47**, 12727 (1993).

[3] B. Qiu, Z. Tian, A. Vallabhaneni, B. Liao, J.M. Mendoza, O.D. Restrepo, Xiulin Ruan, and G. Chen, EPL Europhys. Lett. **109**, 57006 (2015).
[4] D. Li, Y. Wu, P. Kim, L. Shi, P. Yang, and A. Majumdar, Appl. Phys. Lett. **83**, 2934 (2003).
[5] A.I. Boukai, Y. Bunimovich, J. Tahir-Kheli, J.-K. Yu, W.A. Goddard Iii, and J.R. Heath, Nature **451**, 168 (2008).
[6] A.I. Hochbaum, R. Chen, R.D. Delgado, W. Liang, E.C. Garnett, M. Najarian, A. Majumdar, and P. Yang, Nature **451**, 163 (2008).
[7] T. Akiyama, T. Komoda, K. Nakamura, and T. Ito, Phys. Rev. Appl. **8**, 024014 (2017).
[8] A. Seko, A. Togo, H. Hayashi, K. Tsuda, L. Chaput, and I. Tanaka, Phys. Rev. Lett. **115**, 205901 (2015).
[9] T. Markussen, A.-P. Jauho, and M. Brandbyge, Nano Lett. **8**, 3771 (2008).
[10] W. Li, N. Mingo, L. Lindsay, D.A. Broido, D.A. Stewart, and N.A. Katcho, Phys. Rev. B **85**, 195436 (2012).
[11] K. Termentzidis, T. Barreteau, Y. Ni, S. Merabia, X. Zianni, Y. Chalopin, P. Chantrenne, and S. Volz, Phys. Rev. B **87**, 125410 (2013).
[12] S.G. Volz and G. Chen, Appl. Phys. Lett. **75**, 2056 (1999).
[13] D. Donadio and G. Galli, Phys. Rev. Lett. **102**, 195901 (2009).
[14] L. Liu and X. Chen, J. Appl. Phys. **107**, 033501 (2010).
[15] B. Qiu, L. Sun, and X. Ruan, Phys. Rev. B **83**, 035312 (2011).
[16] O. Madelung, U. Rössler, and M. Schulz, Non-Tetrahedrally Bond. Elem. Bin. Compd. I (2002).
[17] J. Callaway, Phys. Rev. **113**, 1046 (1959).
[18] M.G. Holland, Phys. Rev. **132**, 2461 (1963).
[19] N. Mingo, Phys. Rev. B **68**, 113308 (2003).
[20] M. Kazan, G. Guisbiers, S. Pereira, M.R. Correia, P. Masri, A. Bruyant, S. Volz, and P. Royer, J. Appl. Phys. **107**, 083503 (2010).
[21] T.T.T. Nghiêm, J. Saint-Martin, and P. Dollfus, J. Comput. Electron. **15**, 3 (2016).
[22] C. Jacoboni and P. Lugli, *The Monte Carlo Method for Semiconductor Device Simulation* (Springer Vienna, 2012).
[23] H. Hamzeh and F. Aniel, J. Appl. Phys. **109**, 063511 (2011).
[24] S. Mazumder and A. Majumdar, J. Heat Transf. **123**, 749 (2001).
[25] J.-P.M. Péraud and N.G. Hadjiconstantinou, Phys. Rev. B **84**, 205331 (2011).
[26] T. Klitsner, J.E. VanCleve, H.E. Fischer, and R.O. Pohl, Phys. Rev. B **38**, 7576 (1988).
[27] R.B. Peterson, J. Heat Transf. **116**, 815 (1994).
[28] S. Wolf, N. Neophytou, and H. Kosina, J. Appl. Phys. **115**, 204306 (2014).
[29] J. Maire, R. Anufriev, R. Yanagisawa, A. Ramiere, S. Volz, and M. Nomura, Sci. Adv. **3**, e1700027 (2017).
[30] Y. Chen, D. Li, J.R. Lukes, and A. Majumdar, J. Heat Transf. **127**, 1129 (2005).
[31] D. Lacroix, K. Joulain, and D. Lemonnier, Phys. Rev. B **72**, 064305 (2005).
[32] A.L. Moore, S.K. Saha, R.S. Prasher, and L. Shi, Appl. Phys. Lett. **93**, 083112 (2008).
[33] E.B. Ramayya, L.N. Maurer, A.H. Davoody, and I. Knezevic, Phys. Rev. B **86**, 115328 (2012).
[34] H.B.G. Casimir, Physica **5**, 495 (1938).
[35] R. Berman, F.E. Simon, and J.M. Ziman, Proc. R. Soc. Math. Phys. Eng. Sci. **220**, 171 (1953).
[36] J.M. Ziman, *Electrons and Phonons: The Theory of Transport Phenomena in Solids* (Oxford University Press, Oxford, New York, 2001).
[37] D.H. Santamore and M.C. Cross, Phys. Rev. B **63**, 184306 (2001).
[38] S.B. Soffer, J. Appl. Phys. **38**, 1710 (1967).
[39] L. Chaput, J. Larroque, P. Dollfus, J. Saint-Martin, and D. Lacroix, Appl. Phys. Lett. **112**, 033104 (2018).
[40] D. Lacroix, K. Joulain, D. Terris, and D. Lemonnier, Appl. Phys. Lett. **89**, 103104 (2006).
[41] J. Larroque, P. Dollfus, and J. Saint-Martin, J. Phys. Conf. Ser. **906**, 012007 (2017).
[42] A. Togo, L. Chaput, and I. Tanaka, Phys. Rev. B **91**, 094306 (2015).
[43] See Supplementary Materials at [URL].
[44] W. Setyawan and S. Curtarolo, Comput. Mater. Sci. **49**, 299 (2010).
[45] C. Jacoboni and P. Lugli, *The Monte Carlo Method for Semiconductor Device Simulation* (Springer Vienna, Vienna, 1989).
[46] T. Ruf, R.W. Henn, M. Asen-Palmer, E. Gmelin, M. Cardona, H.-J. Pohl, G.G. Devyatych, and P.G. Sennikov, Solid State Commun. **115**, 243 (2000).
[47] Y.S. Ju and K.E. Goodson, Appl. Phys. Lett. **74**, 3005 (1999).
[48] W. Liu and M. Asheghi, J. Heat Transf. **128**, 75 (2006).